\documentclass[aps,prb,reprint,showpacs,superscriptaddress,amsmath,amssymb,amsfonts]{revtex4-1} 

\usepackage{graphicx}
\usepackage{color}
\usepackage{subfigure}
\usepackage{array}

\newcolumntype{M}[1]{>{\centering\arraybackslash}m{#1}}

\begin{document}

\title{Bound states in the continuum in symmetric and asymmetric photonic crystal slabs}

\author{Anton I. Ovcharenko}
\affiliation{Laboratoire Charles Fabry, Institut d'Optique Graduate School, CNRS, Universit\'e Paris-Saclay, 91127 Palaiseau, France}
\author{C\'edric Blanchard}
\affiliation{CEMHTI, CNRS UPR 3079, Orl\'eans University, Orl\'eans, France}
\author{Jean-Paul Hugonin}
\affiliation{Laboratoire Charles Fabry, Institut d'Optique Graduate School, CNRS, Universit\'e Paris-Saclay, 91127 Palaiseau, France}
\author{Christophe Sauvan}\email[Corresponding author: ]{christophe.sauvan@institutoptique.fr}
\affiliation{Laboratoire Charles Fabry, Institut d'Optique Graduate School, CNRS, Universit\'e Paris-Saclay, 91127 Palaiseau, France}

\date{\today}

\begin{abstract}
We develop a semi-analytical model to describe bound states in the continuum (BICs) in photonic crystal slabs. We model leaky modes supported by photonic crystal slabs as a transverse Fabry-Perot resonance composed of a few propagative Bloch waves bouncing back and forth vertically inside the slab. This multimode Fabry-Perot model accurately predicts the existence of BICs and their positions in the parameter space. We show that, regardless of the slab thickness, BICs cannot exist below a cut-off frequency, which is related to the existence of the second-order Bloch wave in the photonic crystal. Thanks to the semi-analyticity of the model, we investigate the dynamics of BICs with the slab thickness in symmetric and asymmetric photonic crystal slabs. We evidence that the symmetry-protected BICs that exist in symmetric structures at the $\Gamma$-point of the dispersion diagram can still exist when the horizontal mirror symmetry is broken, but only for particular values of the slab thickness.
\end{abstract}


\maketitle

\section{Introduction}

Photonic crystals (PhCs) consist of a periodic modulation of the refractive index at the wavelength scale~\cite{JoannopoulosBook,SakodaBook} and PhC slabs are formed by etching a one- or two-dimensional (1D or 2D) PhC in a dielectric layer acting as an optical waveguide. Compared to three-dimensional (3D) PhCs, the simplified architecture of PhC slabs makes them attractive for on-chip integrated photonics \cite{NotomiReview,Kuramochi2016}. In addition, their peculiar diffraction properties have enabled a wide variety of applications, including filters\cite{baets2001optical, Mateus2004, Ding2004}, vertical-cavity surface-emitting lasers (VCSEL)\cite{VCSEL-Sciancalepore}, thermal emission\cite{Inoue2014Termal}, and structural color generation \cite{Brongersma2014,Shen2015}.

It was recently pointed out that PhC slabs can support optical bound states in the continuum (BICs)~\cite{Hsu2013,Blanchard2014,Hsu2016}. A BIC (also called embedded eigenvalue \cite{MonticoneBIC}) is a bound state that exists at the same energy level as a continuum of radiation modes~\cite{vnw1929,Friedrich1985}. In PhC slabs, it corresponds to an eigenmode that is truly guided (no radiative leakage) despite the fact that it lies above the light cone in the dispersion diagram $\omega = f(\mathbf{k})$, with $\omega$ the angular frequency and $\mathbf{k}$ the wavevector. The absence of leakage originates from two different physical mechanisms: a symmetry incompatibility or a destructive interference between different leakage channels~\cite{Hsu2013}.


From a strictly theoretical point of view, BICs, and especially the ones resulting from an interference mechanism, are definitely counter-intuitive and intriguing modes. From a practical point of view, however, BICs do not really exist. Indeed, in a real non-ideal structure, they are anyway faintly coupled to the radiation continuum because of technological imperfections, roughness, or finite size of the device. A BIC thereby becomes a leaky mode with extremely low leakage, i.e., with a very large quality factor, $Q$. Therefore, if a PhC slab can be fabricated with geometrical parameters close enough to the ideal ones, it exhibits a very sharp resonance with an extremely high quality factor whose value is only limited by technological constraints. Such high-$Q$ resonances that result from the existence of a BIC nearby in the parameter space have been recently investigated~\cite{MagnussonSR15,Blanchard2016,SadrievaLavrinenkoACS2017}  and exploited for lasing~\cite{Kodigala2017Nature,Kuznetsov2018} and sensing applications \cite{SensingLiu,SensingRomano,Yesilkoy2019Sensing}.

Up to now, the existence of BICs and their location in the parameter space has been calculated either with rigorous numerical methods \cite{Azzam2018,Bulgakov2018a,Li2016,Wang2016}, coupled-wave theory~\cite{Wang2016a,Yang2014,Ni2016}, or a perturbation approach\cite{Blanchard2014, Yuan2017}. Some interesting proposals of numerical approaches especially suitable for such problems have been put forward\cite{Gao2016,Bulgakov2018b}. Fully numerical approaches are cumbersome even for simple geometries since the whole parameter space has to be explored blindly to find a BIC. Using a perturbation approach is an interesting alternative. However, if the coupled-wave formalism is accurate for PhC slabs with a low refractive-index contrast, the accuracy drops as the contrast increases. Iterative schemes have been proposed to improve the accuracy of the coupled-wave formalism for high refractive-index contrasts but at the cost of a drastic loss in simplicity~\cite{Wang2016a}.

To ease the practical implementation of PhC slabs supporting BICs, one needs approximate models that yield fast yet accurate predictions of the BIC location in the parameter space. Improving the understanding of the physical mechanisms that lead to the BIC formation is also an important issue. We propose a semi-analytical model that does not rely on a perturbative approach. The model presents several advantages. First, it explicitly contains the interference mechanism that leads to the formation of a BIC. Secondly, it yields quantitative predictions of the corresponding optogeometric parameters for any refractive index contrast.

BICs in PhC slabs are leaky modes whose radiative leakage vanishes for a particular set of optogeometric parameters. Their modeling is intrinsically linked to the phenomenon of guided-mode resonance, which corresponds to the resonant excitation of a leaky mode. Over the years, guided-mode resonance has been described by several theoretical formalisms, such as, for instance, coupled-wave theory~\cite{FanJOSA03}, perturbation methods~\cite{Sentenac03}, or polology framework~\cite{Popov86,Sentenac02}. Of a particular interest for the purposes of this work is an approach proposed in 2006 that consists of modeling the reflection and transmission of a PhC slab as a \emph{transverse} Fabry-Perot resonance composed of several waves bouncing back and forth inside the slab~\cite{Lalanne2006}. This multimode Fabry-Perot model has been used with $N=2$ waves to study the optical properties of high contrast gratings (HCGs)~\cite{Lalanne2006,Karagodsky2010,Karagodsky2011,Karagodsky2012}. In this article, we extend the model to $N=3$ waves and apply it to the calculation of the dispersion curve and the quality factor of leaky modes supported by 1D PhC slabs. We show  that the model is able to quantitatively predict the appearance of BICs and their position in the parameter space.

Let us emphasize that the multimode Fabry-Perot model is particularly well-suited for the study of BICs. First, the waves used in the model to build the transverse resonance are exactly the waves that destructively interfere to form a BIC. They have thus a clear physical meaning and are not virtual intermediary means for the calculation, even in the case of structures far from the perturbation regime. Secondly, these waves are bouncing back and forth \emph{vertically} inside the PhC slab. The slab thickness is thus a crucial parameter to understand the formation of BICs by destructive interference. The model predictions are analytical with respect to this geometrical parameter. Thirdly, these waves possess cut-off frequencies below which they cannot propagate. Since the multimode Fabry-Perot model explicitly contains these cut-off frequencies, the zones in the $(\omega, {\bf k})$ space where BICs of different composition, judging by the number of constituent waves, can (or cannot) exist become apparent.

We apply the multimode Fabry-Perot model to 1D PhC slabs that are either symmetric or asymmetric with respect to a vertical axis ($z$-axis in this work), see Fig.~\ref{model} The existence of BICs in asymmetric structures is scarcely documented in the literature \cite{Wang2016, Koshelev2018}. The model by its nature does not discriminate between symmetric or asymmetric slabs; the physical interpretation and the computational cost remains the same as symmetry is broken. Thus, we still benefit from the semi-analytical character of the model and we can quickly explore the parameter space. We evidence the existence of BICs in 1D PhC slabs with a horizontal asymmetry and we study their evolution as the asymmetry parameter is tuned continuously.

We first calculate and describe in Section II the different types of BICs that can exist in a symmetric 1D PhC slab, see Fig.~\ref{model}(a), to set the foundations for further discussion. These different BICs have already been discussed separately in the literature, but here we show that all of them can exist together in a given geometry. In Section III, we present the multimode Fabry-Perot model and we show that it can accurately predict the existence of all types of BICs. We also discuss some limitations of the model. We finally apply the model in Section IV to an asymmetric 1D PhC slab, see Fig.~\ref{model}(b). Section V concludes the work.

\section{Symmetric one-dimensional photonic crystal slabs}

\begin{figure}[t!]
	\includegraphics[width=1\linewidth]{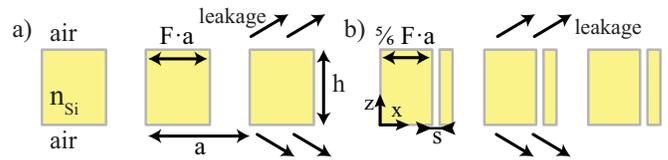}
	\caption{Schematics of both systems under study: a symmetric (a) and an asymmetric (b) 1D PhC slab. Main parameters are the PhC period $a$, the filling factor $\mathrm{F}$, defined as the fraction of dielectric material inside one period, the refractive index $n_d$, and the slab thickness $h$. For the asymmetric PhC slab, a slit of width $s$ is introduced as the asymmetry parameter (no asymmetry for $s=0$).}
	\label{model}
\end{figure}

We consider here a symmetric lamellar 1D PhC slab, that is a periodic array of slits in a dielectric membrane with refractive index $n_d=3.5$ embedded in air, as shown in Fig.~\ref{model}(a). The PhC period, the membrane thickness, and the filling factor in dielectric material are respectively denoted with $a$, $h$, and $F$. We study the leaky modes supported by this structure in Transverse Electric (TE) polarization, i.e., with an electric field polarized along the slits in the $y$ direction.

The modes of the PhC slab are characterized by a wavevector ${\bf k} = (k_x,k_y)$ and an eigenfrequency $\tilde{\omega} = 2\pi c / \tilde{\lambda}$. Because of radiative leakage --~for modes with a real wavevector located above the light cone~-- the eigenfrequencies are complex with a non-zero imaginary part. The latter is related to the mode quality factor, $Q = \mathrm{Re}(\tilde{\lambda})/[2\mathrm{Im}(\tilde{\lambda})]$. Numerical calculations in this work are performed with the rigorous coupled-wave analysis (RCWA) \cite{Moharam1995a}. The modes are calculated by searching for the poles of the scattering matrix in the complex frequency plane~\cite{Bai2013,QNMSolver}. The number of Fourier harmonics retained in the expansion of the electromagnetic field is $2M+1$ with $M = 30$.

\begin{figure*}[htb]
	\centering
	\includegraphics[width=\linewidth]{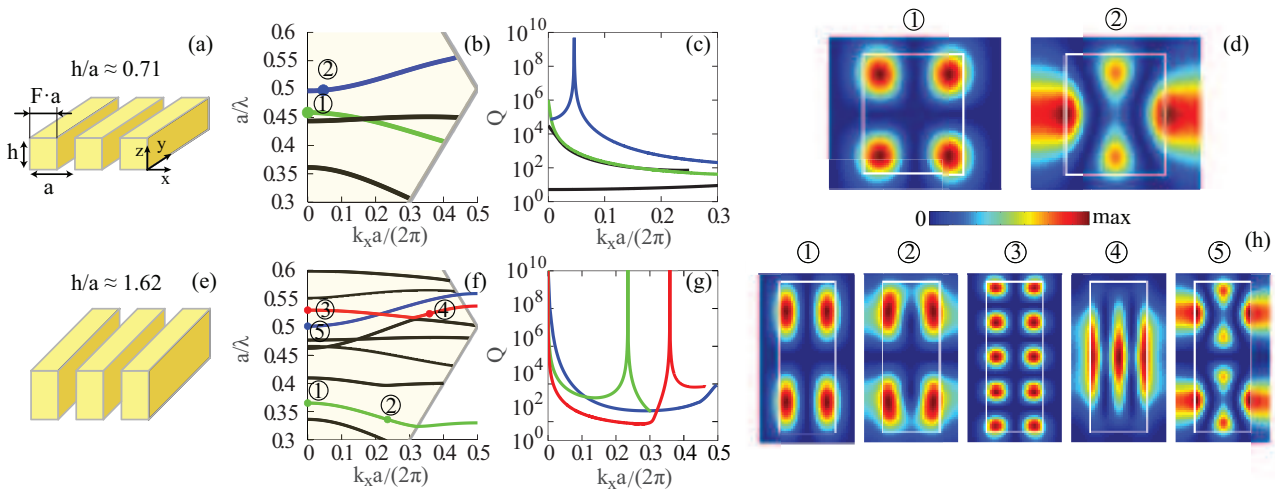}	
 	\caption{BICs in symmetric PhC slabs for two different values of the slab thickness, $h = 0.71a$ (a)-(d) and $h = 1.62a$ (e)-(h). (a) and (e) Schematics of the structures. The filling factor and the refractive index of the dielectric material are fixed, $\mathrm{F} = 0.6$ and $n_d=3.5$. (b) and (f) Dispersion diagrams of the leaky modes above the light cone (gray  line). The bands represented with colored curves exhibit BICs for some particular values of the wave vector $k_x$ shown by colored dots. (c) and (g) Quality factors $Q$ of the leaky modes with the same colors as the dispersion diagram. BICs correspond to $Q$-factors that tend to infinity (numerically larger than $10^9$). (d) and (h) Electric-field distributions $|E_y(x,z)|$ of the BICs shown by colored dots in (b) and (e). A single period is represented; the edges of the dielectric material are shown with white lines.}
 	\label{bics}
 \end{figure*}

Figure~\ref{bics} shows the different types of BICs that can exist in a symmetric 1D PhC slab. It also evidences the crucial role of the slab thickness in the formation mechanism of BICs. Figure~\ref{bics}(b) displays the dispersion curves of the first four leaky modes with the lowest frequency for $h = 0.71a$ and $F = 0.6$. The normalized frequency $a/\mathrm{Re}(\tilde{\lambda})$ of the modes has been calculated as a function of the normalized $x$-component of the wavevector $k_xa/(2\pi)$, which is varied inside the first Brillouin zone, for a fixed $k_y = 0$ (non-conical mount).
The quality factors of the four modes are shown in Fig.~\ref{bics}(c). Two modes (green and blue curves) exhibit a BIC along their dispersion curve while the other two (black curves) do not. Indeed, the quality factor of the green mode diverges for $k_x = 0$ and that of the blue mode diverges for $k_x = 0.046 (2\pi/a)$. The locations of these two BICs in the dispersion diagram are shown with dots labeled (1) and (2) in Fig.~\ref{bics}(b). The corresponding electric fields are displayed in Fig.~\ref{bics}(d).

The existence of the BIC labeled (1) at $k_x = 0$ (the $\Gamma$-point) can be easily understood: radiative leakage is prohibited due to symmetry incompatibility. The field profile of the mode is antisymmetric with respect to $x$, $E_y(-x) = -E_y(x)$, and cannot couple to the symmetric profile of a plane wave with $k_x = 0$. This kind of BICs had been identified in earlier works on PhC slabs; in recent literature they are sometimes referred to as symmetry-protected BICs~\cite{Hsu2013, Hsu2016}.

The existence of the second BIC labeled (2) is, however, more intriguing. Since its field presents no particular symmetry, it should, in principle, be coupled to the radiation continuum. However, radiative leakage is exactly suppressed at $k_x = 0.046 (2\pi/a)$. This ``accidental'' disappearance of leakage results from destructive interference between several leakage channels~\cite{Hsu2013}. In the literature this type of BIC can be referred to as a resonance-trapped \cite{Kodigala2017Nature,Hsu2016}, or a Friedrich-Wintgen BIC \cite{Friedrich1985}.

As the thickness $h$ of the PhC slab increases to $h = 1.62a$ (same filling factor $F = 0.6$), more modes appear in the spectral range of interest and the number of BICs increases as well, see Figs.~\ref{bics}(e)-(h). Our calculations show five BICs whose locations in the dispersion diagram of Fig.~\ref{bics}(f) are marked with the dots labeled from (1) to (5); the corresponding electric fields are shown in Fig.~\ref{bics}(h). Figure~\ref{bics}(g) displays the quality factors of the three modes that exhibit one or two BICs along their dispersion curve. The green mode has a diverging quality factor at $k_x = 0$, a symmetry-protected BIC [antisymmetric field profile (1)], and another one at $k_x = 0.235 (2\pi/a)$ due to destructive interference [field profile (2) with no particular symmetry]. Similarly, the red mode has the same kinds of BICs at $k_x = 0$ [antisymmetric field profile (3)] and $k_x = 0.3587 (2\pi/a)$ [field profile (4)]. Finally, the blue mode at $k_x = 0$ exhibits a BIC despite its symmetric field profile (5) thanks to a destructive interference mechanism. Using the model described in Section III, the difference between BICs (2) and (4) will become apparent: the former can be described by a Fabry-Perot model with $N=2$ waves, while the latter requires $N=3$ waves.


As summarized in Table~1, we can distinguish two different types of BICs in a symmetric 1D PhC slab.
First, leakage can be forbidden because of symmetry incompatibility between the mode of the PhC slab and the plane waves of the radiation continuum. Such symmetry-protected BICs can only exist at $k_x = 0$; they have an antisymmetric field profile. Their existence does not depend on the geometrical parameters of the PhC slab, provided that the horizontal symmetry is conserved.
Secondly, leakage can also be suppressed by destructive interference mechanisms. This can happen equally at $k_x = 0$ (with a symmetric field profile) and at $k_x \neq 0$. In the latter case, the field profile presents no strict symmetry, but it can be sorted in two categories: the field is either quasi-symmetric [e.g., BIC (4) in Fig.~1(h)] or quasi-antisymmetric [e.g., BIC (2) in Fig.~1(h)]. By the prefix quasi, we mean that the BIC belongs to the dispersion curve of a leaky mode that is either symmetric or antisymmetric at $k_x = 0$. In contrast to the first type of BIC, the existence of such BICs formed by destructive interference strongly depends on the geometrical parameters of the PhC slab. It is thus not straightforward to predict their precise position along the dispersion curve.

\begin{table}[h]
\caption{Symmetry properties along the horizontal $x$ axis of the field profile $E_y(x)$ of BICs in symmetric 1D PhC slabs. The BICs that exist due to symmetry incompatibility necessarily have an antisymmetric field profile at $k_x = 0$ and cannot exist at $k_x \neq 0$. The BICs formed by destructive interference can exist equally at $k_x = 0$ (with a symmetric field profile) or at $k_x \neq 0$. An antisymmetric (resp. symmetric) BIC corresponds to $E_y(-x) = -E_y(x)$ [resp. $E_y(-x) = E_y(x)$]. By quasi-symmetric (resp. quasi-antisymmetric), we mean that the field profile of the BIC is almost symmetric (resp. almost antisymmetric), see BICs labeled 2 and 4 in Fig.~\ref{bics}(h).}
\begin{ruledtabular}
\begin{tabular}{c|c|c}

   & Symmetry incompatibility & Destructive interference \\ \hline
  $k_x = 0$ & antisymmetric BIC & symmetric BIC \\ \hline
  $k_x \neq 0$ & no BIC & \begin{tabular}{c} quasi-symmetric BIC or \\ quasi-antisymmetric BIC \\ \end{tabular}

\end{tabular}
\end{ruledtabular}
\end{table}

If previous works have qualitatively explained the destructive interference mechanism that leads to the formation of a BIC, see for instance Ref.~\onlinecite{Hsu2013}, to our knowledge, none has made the argument quantitative. In the following Section, we present a multimode Fabry-Perot model that allows for a simple yet quantitative analysis of the interference mechanism between the different Bloch waves composing a leaky mode. Thanks to its semi-analytical character, the model allows for easy calculations of the BIC positions in the dispersion diagram and their variation as a function of the slab thickness.

\section{Multimode Fabry-Perot model}

We derive a semi-analytical model that predicts the dispersion curve and the quality factor of leaky modes supported by a PhC slab. We extend an approach proposed in Ref.~\onlinecite{Lalanne2006} for the calculation of the reflection and transmission of a PhC slab. A leaky mode is nothing but a transverse Fabry-Perot resonance composed of several Bloch waves (BWs) bouncing back and forth \emph{vertically} inside the slab. This description is perfectly rigorous as long as a sufficiently large number $M$ of waves is taken into account. This is the mathematical ground of RCWA, also known as the Fourier modal method \cite{Moharam1995a}. In the case of subwavelength periodic structures, only a small number $N$ of BWs are propagative, the other ones being evanescent~\cite{Lalanne1999,Lalanne2006}. Neglecting the impact of the evanescent waves provides approximate results that can be very accurate, provided that the slab thickness is large enough, typically larger than the decay length of the least attenuated evanescent wave~\cite{Lalanne2006}.

In the following, we first introduce some notations and write the general equations that lead to an exact calculation of the leaky modes dispersion. We then neglect the evanescent BWs and derive closed-form expressions for the dispersion curve and the quality factor with $N = 1$, $N = 2$, and $N = 3$ propagative waves. We evidence that BICs can only exist when at least two Bloch waves are propagative. We finally validate the model by comparing its semi-analytical predictions to the exact calculations shown in Fig.~\ref{bics}.

\begin{figure*}[htb]
	\centering
	\includegraphics[width=\linewidth]{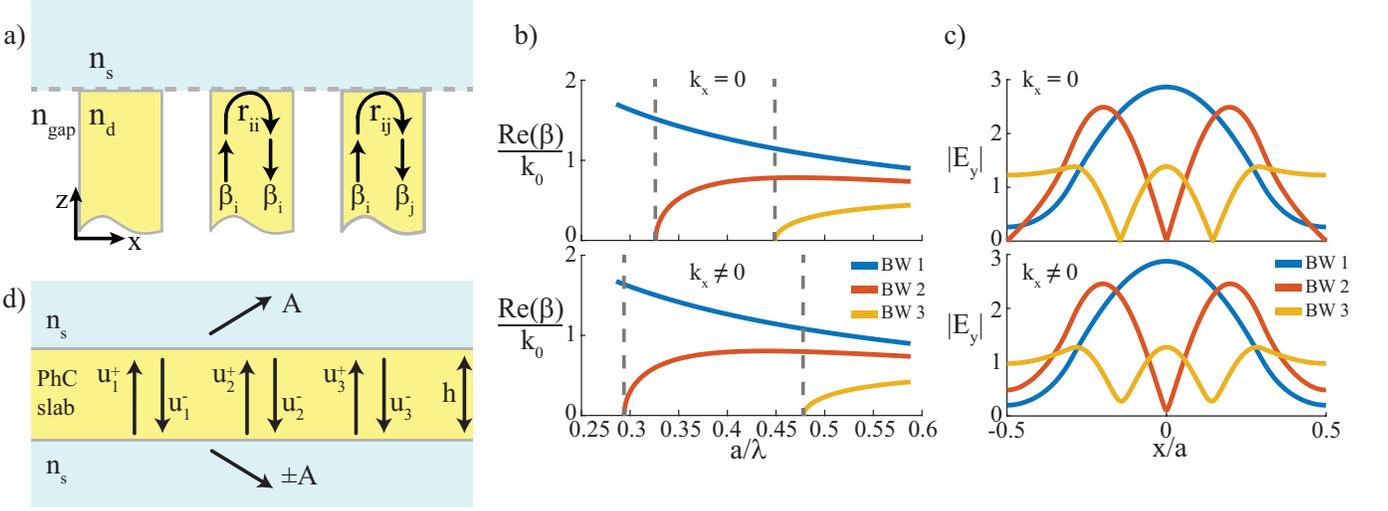}	
	\caption{Multimode Fabry-Perot model. (a) Interface between a semi-infinite PhC and a homogeneous medium. The refractive index of the homogeneous medium $n_s$ can be different from the index $n_\mathrm{gap}$ inside the slits of the PhC. A the interface, the Bloch waves (propagation constants $\beta_i$) propagating in the periodic medium are reflected with a reflection coefficient $r_{ii}$, cross-reflected with a reflection coefficient $r_{ij}$, or transmitted with a transmission coefficient $t_i$. (b) Normalized propagation constants of the BWs as a function of the frequency at the $\Gamma$-point ($k_x=0$) and at $k_x = 0.2a/(2\pi)$. The cut-off frequencies of the second and third BWs are shown by vertical dashed lines. (c) Electric-field profile $|E_y(x)|$ of the three propagative BWs. (d) Principle of the multimode Fabry-Perot model. In the spectral range of interest, up to three BWs can propagate back and forth inside the PhC slab, all other BWs being evanescent. Each BW is transmitted in the surrounding medium with its own phase. The amplitude $A$ of the plane wave propagating away (radiative leakage) results from the interference between these three contributions.}
	\label{bloch}
\end{figure*}

\subsection{Notations and general equations}

Before building a multimode Fabry-Perot resonance in a PhC slab of thickness $h$, we need to solve the problem of a single interface between a semi-infinite PhC and a semi-infinite homogeneous medium, see Fig. \ref{bloch}(a). We denote by $\beta_i$ the propagation constant of the $i^{\rm th}$ BW along the vertical $z$ direction. In a non-absorbing PhC, $\beta_i$ is either purely real (propagative wave) or purely imaginary (evanescent wave). The number of propagative BWs depends on the geometry. For example, for $k_x=0$, $F=0.6$, and $n=3.5$ in the band $ a/\lambda \in [0.25, 0.6] $, only up to three BWs are propagative in a symmetric lamellar 1D PhC slab. Their propagation constants are shown in Fig.~\ref{bloch}(b), where we can observe the second and third BW cut-offs at $a/\lambda = 0.327$ and $a/\lambda = 0.45$, respectively. Note that the fundamental BW (largest propagation constant, blue curve) has no cut-off and is propagative whatever the $a/\lambda$ ratio. The corresponding field profiles $E_y(x)$ are shown in Fig.~\ref{bloch}(c). The fundamental BW is symmetric and the higher-order BWs have alternately an antisymmetric or a symmetric field profile. For $k_x \neq 0$, the cut-off frequencies vary with $k_x$ and the BWs are no longer strictly symmetric nor antisymmetric.

As the $i^{\rm th}$ BW is incident on an interface with a homogeneous medium, it is reflected with a reflection coefficient $r_{ii}$. In addition, it is reflected into a different BW with a cross-reflection coefficient $r_{ij}$, and transmitted as a propagative plane wave with a transmission coefficient $t_i$. The coefficients $r_{ii}$, $r_{ij}$, and $t_i$ are the generalized Fresnel coefficients for an interface between a homogeneous and a periodic media. Note that we limit ourselves to the case where a single plane wave is propagative in the homogeneous medium --~the zeroth diffraction order of the PhC slab. The energy contained in this plane wave corresponds to the radiative leakage.

In a PhC slab of thickness $h$, the BWs are reflected at the top and bottom interfaces. Thus, they propagate back and forth inside the slab as illustrated in Fig.~\ref{bloch}(d). We denote by $u_i^{+}$ and $u_i^{-}$ the amplitudes of the up- and down-propagating $i^{\rm th}$ wave, respectively. The phase origin for the amplitude $u_i^{+}$ (resp. $u_i^{-}$) is taken at the bottom interface (resp. the top interface).

Leaky modes of a PhC slab are solutions of Maxwell's equations in the absence of an incident wave. If one considers a finite number $M$ of BWs (propagative and evanescent BWs), the amplitudes $u_i^{+}$ and $u_i^{-}$ are related by

\begin{eqnarray}\label{eq:NBW}
	\begin{aligned}
		u_i^{+} = \sum_{j=1}^{M} r_{ji} u_j^{-} \exp(i\beta_j h) ,\\
		u_i^{-} = \sum_{j=1}^{M} r_{ji} u_j^{+} \exp(i\beta_j h) .
	\end{aligned}
\end{eqnarray}

\noindent The number $M$ of BWs is equal to the truncation rank of the Fourier series in RCWA~\cite{Moharam1995a}, $M = 30$ in Fig.~\ref{bics}. For the sake of simplicity, we consider a PhC slab surrounded by the same homogeneous medium above and below. The equations can be straightforwardly generalized to the case of two different media (PhC slab lying over a substrate, see Supplemental Material for more details); two different families of reflection coefficients $r^T_{ji}$ and $r^B_{ji}$ have to be considered \cite{Lecamp2005}.

Equations~\eqref{eq:NBW} can be rewritten in a matrix form

\begin{equation}
	\mathbf{R}(k_x,\lambda)\mathbf{U} = 0,
 \end{equation}

\noindent where the vector $\mathbf{U}$ is built with the amplitudes $u_i^{+}$ and $u_i^{-}$, $\mathbf{U} = [u_1^{+},u_1^{-},...,u_M^{+},u_M^{-}]^t$, and the matrix $\mathbf{R}(k_x,\lambda)$ contains all reflections and cross-reflection coefficients. A leaky mode is a non-trivial solution of this linear system of equations; it corresponds to a pair $(k_x,\tilde{\lambda})$ (with $k_x$ a real number and $\tilde{\lambda}$ a complex number) that satisfies

\begin{equation}\label{eq:detN}
    \mathrm{det}\left[\mathbf{R}(k_x,\tilde{\lambda})\right] = 0 ,
\end{equation}

\noindent with $\mathrm{det}$ being the determinant of a matrix. We calculate rigorously with RCWA the parameters of a single interface ($\beta_i$, $r_{ii}$, $r_{ij}$, $t_i$), and  thus the matrix $\mathbf{R}(k_x,\lambda)$, as a function of the wavelength for a fixed value of the wave vector $k_x$. Then, Eq.~\eqref{eq:detN} can be solved, typically with an iterative procedure such as the generic Newton method or a different method using a Pad\'e approximation \cite{Bai2013}, to find the complex wavelength $\tilde{\lambda}$ of the leaky mode. The dispersion curve and the quality factor are then given respectively by $a/\mathrm{Re}(\tilde{\lambda}) = f(k_x)$ and $Q = \mathrm{Re}(\tilde{\lambda})/[2\mathrm{Im}(\tilde{\lambda})]$.

Regarding the radiative leakage, the amplitude of the outgoing propagative plane wave is given by

\begin{equation}\label{eq:AmpN}
    A = \sum_{j=1}^{M} t_j u_j^{+} \exp(i\beta_j h) .
\end{equation}

\noindent The radiative leakage results from the interference of the BWs amplitudes being transmitted by the interface. \emph{Therefore, the leaky mode of the PhC slab is a BIC if, and only if, the interference is perfectly destructive}. One readily realizes the crucial role of the slab thickness $h$ in this mechanism since it drives the value of the phase difference between the different BWs.

Solving Eq.~\eqref{eq:detN} for a large number $M$ of BWs, hence containing a bunch of evanescent waves, yields a rigorous and exact result for the dispersion curve and quality factor. On the other hand, since the period of the PhC slab is subwavelength, neglecting all the evanescent BWs to keep only a small number $N$ of propagative BWs drastically reduces the size of the linear system in Eq.~\eqref{eq:NBW}. Within this approximation, it is possible to derive closed-form expressions for the dispersion curve, the quality factor $Q$, and the radiative leakage $A$, as shown hereafter. In particular, these expressions provide analytical results with respect to the thickness $h$.

\subsection{Transverse resonance for $N=1$ wave}

Let us start with the simplest case $N=1$ when a single BW is propagative inside the PhC slab, all the other waves being evanescent. Although self-evident, this case allows us to introduce the main equations of the model. The single-mode regime occurs when the period-to-wavelength ratio $a/\lambda$ is small, typically between the limit $a/\lambda \rightarrow 0$ (quasi-static limit) and the cut-off of the second BW. For the example in Fig.~\ref{bloch}(b), it corresponds to $a/\lambda < 0.272$ across all $k_x$ from zero to the light line.

For $M=1$ (a single propagative BW), Eq.~\eqref{eq:detN} simply reduces to the usual resonance condition of a Fabry-Perot resonator

\begin{equation}\label{eq:det1}
    1- r_{11}^2\exp(2i\beta_1 h) = 0 \,.
\end{equation}

\noindent Provided that the quality factor of the resonance is large, $Q \gg 1$, and the modulus of $r_{11}$ varies slowly with the wavelength over the resonance bandwidth, closed-form expressions for the phase-matching condition and the quality factor can be derived.

An eigenmode of the PhC slab corresponds to a BW that returns in phase after half a round trip~\cite{HausBook,LPR2008},

\begin{equation}\label{eq:phase1BW}
    \Phi_T(\lambda_0,k_{x}) = \beta_1 h + \mathrm{arg}(r_{11}) = p \pi \,,
\end{equation}

\noindent where $\lambda_0 = \mathrm{Re}(\tilde{\lambda})$ and $p$ is an integer. The phase $\Phi_T$ is the total phase accumulated by the BW after half a round trip inside the slab. This phase-matching condition gives an implicit definition of the dispersion curve.

The quality factor $Q$ is given by~\cite{PRB2005,LPR2008}

\begin{equation}\label{eq:Q1BW}
    Q = - \frac{\lambda_0}{1-|r_{11}|^2} \frac{\partial \Phi_T}{\partial\lambda} \,.
\end{equation}

\noindent Finally, the amplitude $A$ of the radiated plane wave is simply proportional to the BW amplitude inside the slab, $A = t_1 u_1^{+}\exp(i\beta_1 h)$. In this case, the leakage does not result from the interference between several channels; it vanishes if, and only if, the transmission $t_1$ is strictly equal to zero. This, however, never happens for symmetry reasons~\footnote{For Bloch-wave-to-plane-wave transmission $t_i$ to disappear due to symmetry incompatibility, the BW has to have an antisymmetric field profile at $k_x = 0$, in contrast to the symmetric profile of the plane wave. Since the fundamental Bloch wave $i=1$ has a symmetric field profile at $k_x = 0$, $t_1$ has always a non-zero value.}.

The Fabry-Perot model allows us to draw an important conclusion: \emph{no BIC can exist at a frequency where a single BW is propagative}. This result sets a spectral cut-off to this existence of BICs, see Fig.~\ref{bloch}(b). We emphasize that this cut-off is independent of the slab thickness.

\subsection{Transverse resonance for $N=2$ waves}

For larger period-to-wavelength ratios, the second BW, which has an antisymmetric field profile, becomes propagative, see Fig.~\ref{bloch}(b). Let us start with the situation $k_x = 0$. Since the fundamental BW is symmetric whereas the second BW is antisymmetric, see Fig.~\ref{bloch}(c), the cross-reflections $r_{12}$ and $r_{21}$ are equal to zero and Eqs.~\eqref{eq:NBW} reduce to two uncoupled sets of two equations each. As a consequence, leaky modes result from a transverse resonance built either with the fundamental BW alone or with the second BW alone. The dispersion curve and the quality factor are given by Eqs.~\eqref{eq:phase1BW} and~\eqref{eq:Q1BW} with either ($\beta_1$, $r_{11}$) or ($\beta_2$,$r_{22}$).

Because of the symmetry mismatch between the BW and the propagative plane wave, $|r_{22}|=1$ and $t_2=0$. Therefore, the mode of the PhC slab that corresponds to a transverse resonance built with the second BW alone is necessarily a BIC whatever the geometrical parameters. In particular, varying the slab thickness $h$ shifts the dispersion curve according to the phase-matching condition but the $Q$-factor remains infinite ($|r_{22}|=1$) and this mode at $k_x=0$ is truly guided with no radiative leakage. It is the aforementioned symmetry-protected BIC, which results from symmetry incompatibility, see Table 1.

As we depart from the $\Gamma$-point, the BWs begin to couple with each other since $r_{12} \neq 0$ and $r_{21} \neq 0$. As a consequence, Eqs.~\eqref{eq:NBW} become a set of four coupled equations and any transverse resonance results from the interplay between both BWs. It is possible to replace such two-wave Fabry-Perot resonator with the usual single-wave one by introducing an effective reflection coefficient $r_\mathrm{eff}$ \cite{Lecamp2005}. The effective reflection fully includes the impact of the second wave. For $k_x=0$, a leaky mode is a purely single-wave transverse resonance built either with the fundamental BW or with the second BW. When $k_x$ becomes non-zero, both BWs are mixed but one keeps a larger contribution than the other. For instance, for a band with a symmetry-protected BIC at $k_x=0$, $|r_{11}| < |r_{22}|$ and $|r_{12}| < |r_{22}|$. In that case, the second BW is dominant and we incorporate the effect of the first BW in the effective reflection coefficient. The resonance condition given by Eq.~\eqref{eq:det1} becomes

\begin{equation}\label{eq:deteff}
    1 - \left( r_\mathrm{eff}^{(12)} \right)^2 \exp(2i\beta_2 h) = 0 \,,
\end{equation}

\noindent where the effective reflection $r_\mathrm{eff}^{(12)}$ is given by

\begin{equation}\label{eq:reff}
r_\mathrm{eff}^{(12)} = \frac{r_{22} + \alpha r_{11} r_{21} r_{12} \exp( 2i\beta_1 h)}
              {1 -\alpha r_{21} r_{12} \exp\left[ i(\beta_1 + \beta_2)h \right] } \,,
\end{equation}

\noindent with $\alpha = \left [ 1 - r_{11}^2 \exp(2i\beta_1 h) \right ]^{-1}$. The superscript $^{(12)}$ stands for the fact that $r_\mathrm{eff}^{(12)}$ includes the multiple cross-reflections between BWs 1 and 2. Note that for $k_x=0$, since $r_{12}=r_{21}=0$, we recover $r_\mathrm{eff}^{(12)} = r_{22}$. Details on the derivation of Eq.~\eqref{eq:deteff} can be found in the Supplemental Material.

We can thus apply the usual equations of a Fabry-Perot resonator. The dispersion curve and the quality factor of a leaky mode composed of two BWs are given by Eqs.~\eqref{eq:phase1BW} and~\eqref{eq:Q1BW} by replacing $\beta_1$ and $r_{11}$ by $\beta_2$ and $r_\mathrm{eff}^{(12)}$. The amplitude of the radiated plane wave is now given by the superposition of both BWs, $A = t_1 u_1^{+} \exp(i\beta_1 h) + t_2 u_2^{+} \exp(i\beta_2 h)$. Similarly to the resonance condition, an effective transmission coefficient can be introduced, $A = t_\mathrm{eff}^{(12)} u_2^{+} \exp(i\beta_2 h)$, with

\begin{equation}\label{eq:teff}
    t_\mathrm{eff}^{(12)} = t_2 + t_1 \alpha r_{21} e^{i\beta_1 h} \left [ r_\mathrm{eff}^{(12)} e^{i\beta_2 h} + r_{11} e^{i\beta_1 h} \right ] \,.
\end{equation}

\noindent Again, for $k_x=0$, $t_\mathrm{eff}^{(12)} = t_2$ since $r_{21}=0$. Note that Eqs~\eqref{eq:deteff}-\eqref{eq:teff} have been written in the case where the second BW is dominant over the first one. They can also be written in the case where the first BW is dominant over the second one by inverting subscripts 1 and 2.

\begin{figure}[htbp]
	\includegraphics[width=.9\linewidth]{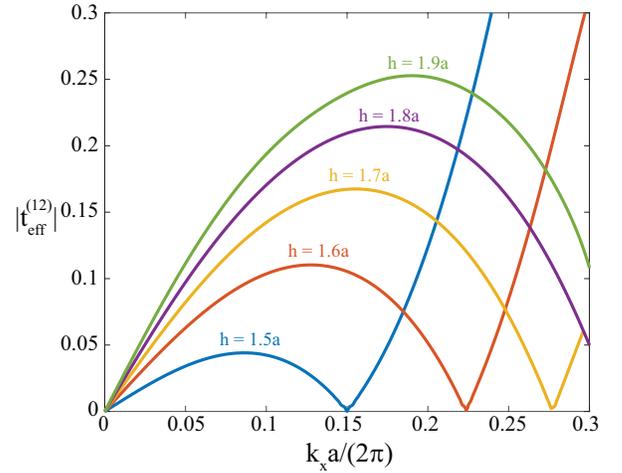}
	\caption{Effective transmission (solid lines) for $N=2$ propagative BWs in a 1D symmetric PhC slab with $F=0.6$. The first cancellation of $t_\mathrm{eff}$ for $k_x=0$ results from symmetry arguments. The second cancellation for $k_x \neq 0$ results from destructive interferences between both BWs and varies with the slab thickness. Dashed lines show the transmission coefficent $t_2$ alone and evidence the impact of the first BW on the radiative leakage.}
	\label{teff}
\end{figure}

One readily realizes that the effective transmission can be canceled if the second term in Eq.~\eqref{eq:teff} is equal to $-t_2$. In that case, both BWs interfere destructively to cancel the overall leakage, leading to the formation of a BIC. Figure~\ref{teff} illustrates the interference mechanism as a function of $k_x$ for different values of the slab thickness. For $k_x = 0$, $t_\mathrm{eff}^{(12)} = t_2 = 0$ for symmetry reasons. The second cancellation of $t_\mathrm{eff}^{(12)}$ is due to destructive interferences. The slab thickness drives the phase difference between both BWs and thus the wave vector value that corresponds to destructive interference increases with $h$.

\subsection{Transverse resonance for $N=3$ waves}

As the period-to-wavelength ratio is further increased, the third BW becomes propagative, see Fig.~\ref{bloch}(b). For $k_x=0$ and $F=0.6$, this corresponds to $a/\lambda > 0.45$. In that case, Eqs.~\eqref{eq:NBW} become a $6 \times 6$ system. Although more tedious, it is still possible to replace the complex interplay between the three BWs by effective reflection and transmission coefficients.

AS for $N=2$, let us start the discussion with the case $k_x=0$. At the $\Gamma$-point, the first and third BWs have a symmetric field profile while the second BW is antisymmetric. The latter is thus decoupled from BWs 1 and 3. Even if three BWs are propagative, the leaky modes of the PhC slab are either formed by the second BW alone (symmetry-protected BIC) or by the interplay between first and third BWs. In that case, we can apply the results from previous section by introducing an effective reflection coefficient $r_\mathrm{eff}^{(13)}$ instead of $r_\mathrm{eff}^{(12)}$. Such a leaky mode formed by BWs 1 and 3 is a BIC if the interference leads to $|r_\mathrm{eff}^{(13)}|=1$ and $t_\mathrm{eff}^{(13)}=0$. This is the case of the blue mode labeled (5) in Figs.~\ref{bics}(f)-(h).

%
%

For $k_x \neq 0$, the three BWs are coupled and we introduce an effective reflection coefficient $r_\mathrm{eff}^{(123)}$, whose closed-form expression can be found in the Supplemental Material. In the process of replacing two BWs by an effective reflection, one keeps the BW that has the most important contribution. The latter is either the second BW or the third BW, depending on the horizontal symmetry of the leaky mode that we want to form.


\subsection{Model validation}

\begin{figure}[htbp]
	\includegraphics[width=1\linewidth]{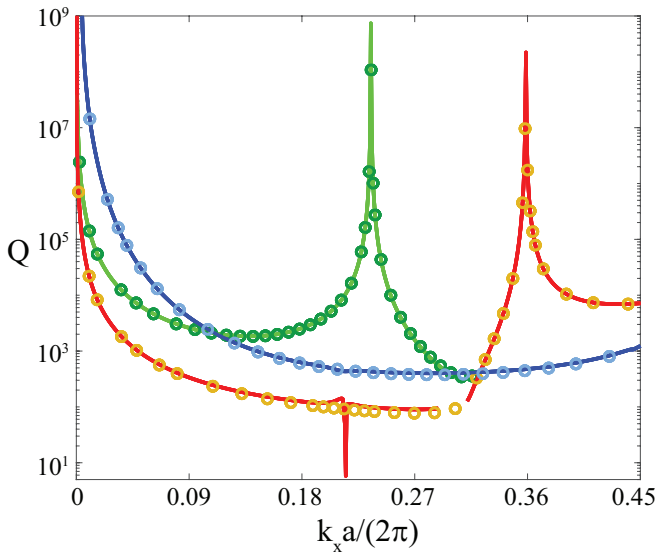}
	\caption{Quality factors of three leaky modes supported by a 1D symmetric PhC slab with $F=0.6$ and $h=1.62a$, see Fig.~\ref{bics}(e). The predictions of the multimode Fabry-Perot model (solid lines) are in excellent agreement with the exact calculations (markers). The model accurately reproduces the existence and positions of all types of BICs.}
	\label{validation}
\end{figure}

To validate the multimode Fabry-Perot model, we consider the symmetric 1D PhC slab of Fig.~\ref{bics}(e). We apply the model, either with $N=2$ or $N=3$ BWs and calculate the dispersion curves and quality factors of the different leaky modes supported by the PhC slab. The model prediction for the dispersion curves (not shown here) and for the Q-factors (see Fig.~\ref{validation}) are in quantitative agreement with the rigorous RCWA calculations, which takes into account a large number of evanescent BWs ($M=30$). In particular, the semi-analytical model accurately predicts all different types of BICs supported by a 1D symmetric PhC slab.

The model also provides some physical insight into the nature of the BICs, which before we could only qualitatively infer from the field profiles. The green mode is mostly given by the second antisymmetric BW, with a small contribution of the fundamental BW for $k_x \neq 0$. To calculate it we used $r_{\mathrm{eff}}^{(12)}$ from Eq.~\eqref{eq:reff}. The blue mode has a quasi-symmetric profile and is dominated by the third BW with a small impact of BWs 1 and 2. For this mode, $r_{\mathrm{eff}}^{(123)}$ with the phase factor $\exp(i\beta_3h)$ was used. We can also assert that the BIC at the $\Gamma$-point is fundamentally different from the other two, being a resonance-trapped BIC resulting from the destructive interference of BWs one and three. It somewhat harder to distinguish from the symmetry protected one with the exact calculations only, and have been pointed out in the literature before \cite{Kodigala2017Nature}. Similarly, the red leaky mode also consists of the three BWs. However, since it couples with a different leaky mode around $k_x \approx 0.3 (2\pi)/a$, it changes its symmetry: from the quasi-antisymmetric one (for $k_x < 0.3 (2\pi)/a$) to the quasi-symmetric one (for $k_x > 0.3 (2\pi)/a$), which is reflected in the BIC field profiles in Fig.~\ref{bics}(h). To correctly apply the model for this case, a slightly different equations are to be used, as was done for the left and right parts of the red curve. Basically, indices $2$ and $3$ are to be interchanged in the expressions for $r_{\mathrm{eff}}^{(123)}$ and the phase matching condition, to reflect the fact that on one side the second BW is dominant, and on the other -- the third one.
This way, $r_{\mathrm{eff}}^{(123)}$ on the left side of the figure becomes equal to $r_{22}$ at $k_x = 0$ and gives rise to the symmetry-protected BIC.
While the other one, at $k_x \approx 0.36 (2\pi)/a$, stems from the interference of the three BWs, the third (symmetric) being the dominant one.

This aspect of the model can be a source of erroneous results. Since we approximate the system, as a resonator with only one wave that has a complex reflection coefficient, phase-matching gives us unperturbed resonances, instead of anti-crossings \cite{Bulgakov2018b, Azzam2018}. This is the reason why the model fails in Fig.~\ref{validation} for the red curve around $k_x \approx 0.3 (2\pi)/a$, where the Rabi splitting occurs and two dispersion curves distance from each other. This point is further illustrated in the Supplemental Material.

Next, we would like now to clarify the obviously erroneous feature of the red curve around $k_x \approx 0.215 (2\pi)/a$. As mentioned above, it is possible to write the phase-matching condition in the from of Eq.~\eqref{eq:phase1BW} under the assumption that the modulus of the reflection coefficient of the mode inside the resonator varies smoothly with the wavelength. This is not, however, necessarily always the case when we deal with effective coefficients. This `spike' corresponds to exactly this situation, when $ |r_{\mathrm{eff}}^{(123)}| $ experiences a sudden resonance-like change, which causes phase-matching to fail (see Supplemental Material for an illustration).

All described BICs are robust in the sense that a slight change of one parameter, for example, increase of the thickness, will cause the symmetry-protected resonance red-shift, according to the dispersion relation. In case of the resonance-trapped BICs, they evolve in the whole $\left( k_x, h, \lambda \right)$ parameter space.

 \begin{figure}[hbtp]
 	\centering
 	\includegraphics[width=.95\linewidth]{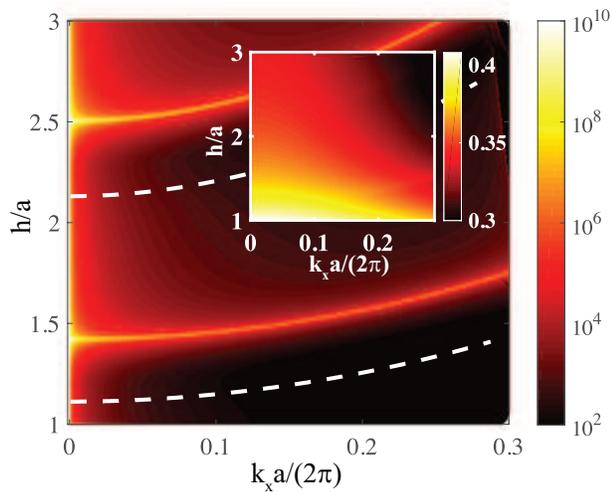}
 	\caption{Quality factor of the mode inside the grating in the logarithmic scale vs. $h/a$ and $k_x a/(2\pi)$. The wavelength range specified during the calculations is such that the only two Bloch waves are propagating. The integer $p$, from Eq.\eqref{eq:phase1BW} is set to $1$, which corresponds to the green line in Fig.~\ref{bics}(f)-(g). Two resonance branches are clearly visible. Dashed white lines indicate the positions of their counterparts for $p=0$ (a different leaky mode). The inset shows the values of $a/\lambda$ (the dispersion) for the same mode and parameter space, as the main figure.}
 	\label{pcolorQ12}
 \end{figure}

\section{Dynamics of BICs with the slab thickness}

Thanks to the analyticity of the model with respect to the slab thickness, we can apply the model for a large number of $h$ values to observe the dynamics of BICs with no additional RCWA calculations. Figure \ref{pcolorQ12} shows the variation of the quality factor of a leaky mode of the PhC slab, in the spectral range between the second and third BW cut-offs, as a function of the thickness $h/a$ and the wave vector $k_x a/(2\pi)$. We have chosen the mode represented with the green curve in Fig.~\ref{bics}(f)-(g). The inset shows its dispersion -- the values of the normalized frequency $a/\lambda$.


 We can clearly see two branches where $r^{(\mathrm{eff})}_{12}$ equals exactly 1, leading to an infinite $Q$-factor. For the $(h,k_x)$ values corresponding to this branch, the radiative leakage disappears because two BWs involved interfere destructively. For $k_x = 0$, the $Q$-factor is infinite for any value of $h$, since $r^{(\mathrm{eff})}_{12} = r_{22} = 1$. It is necessarily a symmetry protected BIC, as the resonance-trapped one is not allowed at $k_x=0$ below the third BW cut-off. For similar figures in the spectral band where three BWs are propagative, we refer the reader to the Supplemental Material.

\begin{figure}[htbp]
	\centering
	\includegraphics[width=.95\linewidth]{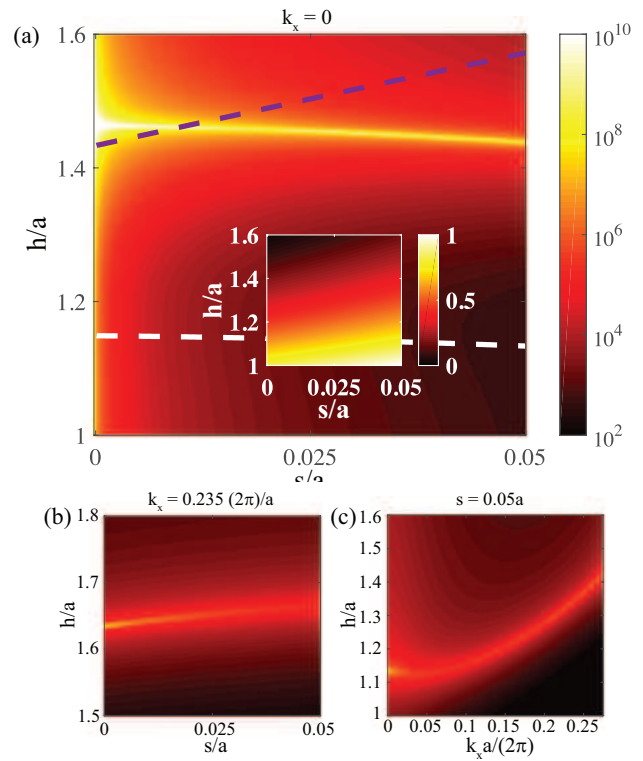}
	\caption{Quality factors for an asymmetric PhC slabs with two propagative BWs. (a) $Q$-factor as a function of the slab thickness $h/a$ and the asymmetry parameter $s/a$ for $k_x=0$ and $p=1$. A BIC (diverging quality factor) is clearly visible whatever the value of the asymmetry parameter. The dashed white line indicates the position of a similar BIC for $p=0$ and the dashed magenta line indicates the position of a BIC formed by three BWs with $p=0$. The inset shows $a/\lambda$ vs. $h/a$ and $s/a$ (the dispersion relation) for the mode of the main figure. (b) Same as (a), but for $k_x \approx 0.235 (2\pi/a)$. (c) $Q$ vs. $h/a$ and $k_x a/ (2\pi)$ for $s=0.05a$. Color scale and other optogeometric parameters are the same for all three subfigures.}
	\label{asym}
\end{figure}

\section{Asymmetric one-dimensional photonic crystal slabs}

We finally use the multimode Fabry-Perot model to study the behavior of the BICs under broken inversion symmetry. We consider a 1D PhC slab with a vertical mirror symmetry but no horizontal mirror symmetry as depicted in Fig.~\ref{model}(b). The asymmetry parameter $s$ is the size of the air gap that divides the dielectric ridge into a bigger and a smaller parts of sizes $5/6Fa$ and $1/6Fa$, respectively, after is was split preserving the $F=0.6$. Recently, a study of a PhC slab with a slot in different positions, with a fixed thickness and normal incidence angle, has been published \cite{Wang2016}. We evidence that the symmetry-protected BICs that exist in symmetric structures at the $\Gamma$-point of the dispersion diagram can still exist when the horizontal mirror symmetry is broken, but only for particular values of the slab thickness.

Figure~\ref{asym} shows a logarithmic map of a $Q$-factor of a leaky mode in different situations. In Fig.~\ref{asym}(a) we plot $Q$ vs. $h$ and $s$ for a fixed $k_x=0$ in a spectral range where only two BWs are propagative. We readily observe a branch that corresponds to a diverging $Q$. Under the broken symmetry, $r_{12}$ no longer vanishes for $k_x=0$, which allows us to observe this trajectory of a resonance-trapped BIC. Figure~\ref{asym}(b) has all the same parameters, except that we now set $k_x$ to a specific non-zero value. Namely, such as to get the BIC labeled (2) in Fig.~\ref{bics}(f) for $s=0$ and see how it disappears very rapidly when the symmetry is broken. Lastly, Fig.~\ref{asym}(c) displays $Q$-factor of the same mode as a function of $h$ and $k_x$ (similarly, as in Fig.~\ref{pcolorQ12}) with a fixed $s = 0.05a$.

\section{Conclusion}

We have used a multimode Fabry-Perot model to calculate the dispersion curves and the quality factors of leaky modes supported by 1D symmetric and asymmetric PhC slabs. Leaky modes are transverse Fabry-Perot resonances composed of a few propagative Bloch waves bouncing back and forth vertically inside the slab. This multimode Fabry-Perot model, which does not rely on a perturbative approach, accurately predicts the existence of BICs and their positions in the parameter space regardless of the refractive index contrast. The model equally applies to symmetry-protected BICs (absence of leakage is due to symmetry incompatibility between a single BW composing the mode and the radiative plane waves) and resonance-trapped BICs (radiative leakage accidentally disappears because the contributions of several BWs interfere destructively).

The multimode Fabry-Perot model allows us to show that, regardless of the slab thickness, BICs cannot exist below a cut-off frequency, which is related to the existence of the second-order Bloch wave in the photonic crystal. In other words, BICs cannot exist in the homogenization regime. Thanks to the semi-analyticity of the model, we investigate the dynamics of BICs with the slab thickness in symmetric and asymmetric photonic crystal slabs. We evidence that the symmetry-protected BICs that exist in symmetric structures at the $\Gamma$-point of the dispersion diagram can still exist when the horizontal mirror symmetry is broken, but only for particular values of the slab thickness.

Since the multimode Fabry-Perot model yields fast yet accurate predictions of the BIC location in the parameter space and provides a better understanding of the physical mechanisms that lead to the BIC formation, we think that it can become an important tool for designing PhC devices relying on the existence of a BIC.

\bibliography{refs}

\end{document}